\documentclass[NoLineNumbers,NewProcedings,InsideFigs]{ascelike-new}
\usepackage[utf8]{inputenc}
\usepackage[T1]{fontenc}
\usepackage{lmodern}
\usepackage{graphicx}
\usepackage[figurename=Fig.,labelfont=bf,labelsep=period]{caption}
\usepackage{subcaption}
\usepackage{amsmath}
\usepackage{longtable}
\usepackage{amsbsy}
\usepackage{newtxtext,newtxmath}
\usepackage[colorlinks=flase]{hyperref}
\usepackage{comment}
\usepackage{upgreek}
\usepackage{multirow}
\usepackage{color}
\usepackage{ulem}
\usepackage{xcolor}

\DeclareRobustCommand{\erase}{\bgroup\markoverwith{\textcolor{black}{\rule[.5ex]{2pt}{0.4pt}}}\ULon}

\begin{document}

	\title{Latent Space-based Likelihood Estimation Using a Single Observation for Bayesian Updating of a Nonlinear Hysteretic Model}

	\author[1]{Sangwon Lee}
	\author[2]{Taro Yaoyama}
	\author[3]{Yuma Matsumoto}
	\author[4]{Takenori Hida}
	\author[5]{Tatsuya Itoi}
	
	\affil[1]{Project researcher, Department of Architecture, Graduate School of Engineering, The University of Tokyo, Tokyo, Japan, Email: lee-sangwon@g.ecc.u-tokyo.ac.jp}
	\affil[2]{Project Assistant Professor, Department of Architecture, Graduate School of Engineering, The University of Tokyo, Tokyo, Japan, Email: yaoyama@g.ecc.u-tokyo.ac.jp}
	\affil[3]{Ph.D student, Department of Architecture, Graduate School of Engineering, The University of Tokyo, Research Fellow of Japan Society for the Promotion of Science, Tokyo, Japan, Email: matsumoto@load.arch.t.u-tokyo.ac.jp}
	\affil[4]{Associate Professor, Major in Urban and Civil Engineering, Graduate School of Science and Engineering, Ibaraki University, Ibaraki, Japan, Email: takenori.hida.mn75@vc.ibaraki.ac.jp}
	\affil[5]{Associate Professor, Department of Architecture, Graduate School of Engineering, The University of Tokyo, Tokyo, Japan, Email: itoi@g.ecc.u-tokyo.ac.jp}
	
	\maketitle

	\begin{abstract}
		This study presents a novel approach to quantify uncertainties in Bayesian model updating, which is effective for sparse or single observations. \color{black} Conventional uncertainty quantification methods are limited in situations with insufficient data, \color{black}particularly for nonlinear responses like post-yield behavior. Our method addresses this challenge \color{black} using \color{black} the latent space of a variational autoencoder (VAE), a generative model that enables nonparametric likelihood evaluation. This approach is valuable in updating model parameters based on nonlinear seismic responses of \color{black} a \color{black} structure, wherein data scarcity is a common challenge. Our numerical experiments confirm the ability of the proposed method to accurately update parameters and quantify uncertainties using a single observation. Additionally, \color{black} the \color{black} numerical experiments reveal \color{black} that \color{black} increased information about nonlinear behavior \color{black} tends \color{black} to result in decreased uncertainty in terms of estimations. This study provides a robust tool for quantifying uncertainty in scenarios characterized by considerable uncertainty, thereby expanding the applicability of approximate Bayesian updating methods in data-constrained environments.
	\end{abstract}

	\section{Introduction}
		Analytical models that describe the behavior of an existing structure require periodic updates to accurately predict \color{black} the realistic performance of a structure throughout its lifetime. \color{black} This need arises from the changing state of \color{black} structures \color{black} caused by factors \color{black} including \color{black} as deterioration or damage due to seismic events. Common practice entails observational data for these updates that reflects the current condition of \color{black} a \color{black} structure.
		
		Numerous conventional methods for such model updating refine the model deterministically to best fit the data. In this field, sensitivity-based approaches \cite{CollinsJ1974,MottersheadJ2011} have received great attention, while \color{black} frequency response methods have been developed \color{black} \cite{Imregun95,Sipple14}.

		These techniques are useful for assessing events including evaluating damage caused by earthquakes, \color{black} particularly \color{black} in well-posed situations wherein the problem is well-defined and the solutions are stable. Although this deterministic approach is effective \color{black} for evaluating \color{black} such existing damage, \color{black} biases may be introduced when it is \color{black} used for future predictions. Intricate systems, such as structural systems, are riddled with uncertainties that possess nondeterministic features \cite{GoldsteinM2016}.
		Precise model-based predictions necessitate the resolution of these uncertainties \cite{VolodinaV2021}.
		Hence, there is a growing interest in probabilistic methods that adequately account for these uncertainties during model updating.
		Amidst these probabilistic methodologies, the Bayesian method is receiving increasing recognition for its effectiveness in dealing with these uncertainties.
		
		The Bayesian approach operates on the principle of updating beliefs based on new evidence.
		Initial or prior probability is established, based upon existing knowledge or beliefs.
		As new data become available, the prior probability is combined with the likelihood of newly observed data.
		This process produces an updated posterior probability which later becomes the new prior for subsequent observations, thereby facilitating a continuous update of beliefs as more data \color{black} are \color{black} obtained.
		This distinct mechanism seamlessly integrates prior knowledge with newly obtained data within the Bayesian methodology.
		
		Within this framework, the likelihood evaluation, which quantifies uncertainty and evaluates the probability of observed data based on a specific hypothesis, is the crucial transition from prior to posterior probabilities.
		This evaluation ensures the accurate updating of posterior probabilities, thereby resulting in more robust and reliable predictions.

		Numerous studies have been conducted on model updating using the Bayesian approach.
		Among these studies, notable approaches \cite{beck98,Yang23,saito13,Saito10} assume a specific distribution form for the likelihood function.
		Although utilizing simple distributions like the normal distribution \color{black} improves computational convenience, \color{black} it is not immune to potentially unrealistic assumptions.
		Approximate Bayesian computation (ABC) \cite{Pritchard99} has been developed for scenarios where computing the full likelihood is challenging \cite{Beaumont02,Beaumont09}.
		\color{black} Utilizing \color{black} the distance between observation and simulation, this approach aims to approximate the posterior distribution of model parameters.
		Several distance-based metrics \cite{bi17}, including approaches that leverage the Bhattacharyya distance \cite{bhattacharyya46}, are employed in ABC for \color{black} stochastic model updating.
		\color{black}
		Stochastic model updating explicitly addresses both aleatory and epistemic uncertainties of model parameters. In many cases, the aim is to update the distribution parameters of the model parameters using multiple observations instead of updating the model parameters themselves.
		\color{black}
		Distance-based ABC has shown promise at stochastic model updating when a substantial amount of data is available \cite{bi19,kitahara22,kitahara21}.
		However, in some cases, obtaining a considerable amount of data, particularly for nonlinear seismic responses of real-world structure, can be a formidable task.
		Accordingly, this study addresses the challenge of parameter estimation in situations where observational data are limited.
		Conventionally, this problem has been approached using \color{black}  deterministic methods \cite{Hinze20} or methods that assume a specific distribution form for the likelihood function \cite{Song18}.
		
		This study presents a novel approach for nonparametric likelihood estimation aimed at quantifying uncertainties in approximate Bayesian model updating \color{black}using a single observation. \color{black}
		The emergence of nonparametric methods that facilitate the nonparametric likelihood estimation of nonlinear hysteretic models with minimal data can revolutionize model updating.
		This advancement would enhance the robustness and generalization capabilities of Bayesian model updating for uncertain future predictions.

	\section{Methodology}
		
		\subsection{Overview of Bayesian Updating for Nonlinear Systems}
			
			\subsubsection{Nonlinear system and feature sample}
				In model updating involving uncertainty quantification, a linear system is often used to model the actual structure. This includes three elements: the model parameter $\boldsymbol{\uptheta}$, output feature $\mathbf{x}$, and the simulator function $h_l(\cdot)$, which are expressed as
				\begin{equation}
					\mathbf{x} = h_l(\boldsymbol{\uptheta})
				\end{equation}
				where $\mathbf{x}=[x_1, x_2,\dots, x_m]^T$; $\boldsymbol{\uptheta} = [\theta_1, \theta_2, \dots, \theta_n]^T$; $m$ and $n$ represent the element size of the output feature $\mathbf{x}$ and the model parameter $\boldsymbol{\uptheta}$, respectively.
				The output feature includes the structural \color{black} properties \color{black} such as natural vibration frequencies or vibration modes, while a linear simulator $h_l(\cdot)$ includes \color{black} a \color{black} method such as seismic linear response analysis.
				
				In the case of model updating of a nonlinear system that models the nonlinear seismic response of an existing structure, the nonlinear system can be characterized \color{black} as: \color{black}
				\begin{equation}
					\mathbf{x} = h(\boldsymbol{\uptheta} \mid u)
				\end{equation}
				\color{black} where \color{black} the nonlinear simulator $h(\cdot)$ is interpreted as the output from the nonlinear seismic response analysis, with $u$ representing the input ground motion. \color{black} An example of an \color{black} output feature from the nonlinear seismic response analysis is a discretized frequency response function of response, which depends on the model parameter $\boldsymbol{\uptheta}$.

				To update the model parameters, the observed features are required.
				Suppose the number of \color{black} observations \color{black} is $N_\text{obs}$, the observed feature sample $\mathbf{X}_\text{obs}\in \mathbb{R}^{m \times N_\text{obs}}$ is expressed \color{black} as: \color{black}
				\begin{equation}
					\label{eq:Xobs}
					\mathbf{X}_\text{obs}=[\mathbf{x}_{1},\mathbf{x}_{2},\dots,\mathbf{x}_{N_\text{obs}}]
				\end{equation}
			
			\subsubsection{Bayes updating}
				The Bayesian updating is performed by evaluating the conditional probabilities of the parameters based on the observed feature sample $p\left(\boldsymbol{\uptheta}\middle| \mathbf{X}_\text{obs}\right)$ expressed \color{black} as: \color{black}
				\begin{equation}
					p\left(\boldsymbol{\uptheta} \middle| \mathbf{X}_\text{obs}\right) = \frac{p\left(\mathbf{X}_\text{obs}\middle|\boldsymbol{\uptheta}\right) p\left(\boldsymbol{\uptheta}\right)}{p\left(\mathbf{X}_\text{obs}\right)}
				\end{equation}
				where
				\begin{itemize}
					\item $p\left(\boldsymbol{\uptheta} \middle| \mathbf{X}_\text{obs}\right)$ is the posterior distribution representing the updated knowledge based on the observational data;
					\item
					$p\left(\mathbf{X}_\text{obs}\middle|\boldsymbol{\uptheta}\right)$ is the likelihood function of $\mathbf{X}_\text{obs}$ for an instance of $\boldsymbol{\uptheta}$;
					\item
					$p(\boldsymbol{\uptheta})$ is a prior distribution representing initial knowledge about $\boldsymbol{\uptheta}$;
					\item
					$p\left(\mathbf{X}_\text{obs}\right)$ is the normalization factor that ensures that the posterior distribution integrates to unity.
				\end{itemize}
				
				Considering that $p\left(\mathbf{X}_\text{obs}\right)$ is a constant,
				the posterior distribution can be expressed \color{black} as: \color{black}
				\begin{equation}
					\label{bayes1}
					p\left(\boldsymbol{\uptheta}\middle|\mathbf{X}_\text{obs}\right)\propto
					p\left(\mathbf{X}_\text{obs}\middle|\boldsymbol{\uptheta}\right)\ p(\boldsymbol{\uptheta})
				\end{equation}
				Using Markov Chain Monte Carlo (MCMC) methods, samples from the posterior $p\left(\boldsymbol{\uptheta}\middle|\mathbf{X}_\text{obs}\right)$ can be obtained.
				Assuming that the prior distribution $p(\boldsymbol{\uptheta})$ is obtained from prior information,
				only the likelihood $p\left(\mathbf{X}_\text{obs}\middle|\boldsymbol{\uptheta}\right)$ must be known from the observation.
				
				Suppose the required size of the samples is $N_\text{sim}$, the simulator $h(\cdot)$ is executed $N_\text{sim}$ times to generate the simulated feature sample
				$\mathbf{X}_\text{sim} \in \mathbb{R}^{m \times N_\text{sim} }$, which can be expressed \color{black} as: \color{black}
				\begin{equation}
					\label{eq:Xsim}
					\mathbf{X}_\text{sim}=[\mathbf{x}_{1},\mathbf{x}_{2}, \dots,\mathbf{x}_{N_\text{sim}}]
				\end{equation}
				
				The number of simulations $N_\text{sim}$ is determined to ensure an accuracy of $p\left(\mathbf{X}_\text{obs}\middle|\boldsymbol{\uptheta}\right)$.
				However, obtaining \color{black} satisfactory \color{black} $N_\text{obs}$ remains \color{black} challenging: the availability of observed data is often limited. \color{black}
				For example, when updating parameters related to the nonlinear structural response during seismic excitation, the number of observations $N_\text{obs}$ is generally few over the lifetime of the structure.
				Such sparse but valuable data can be perceived as outliers in the conventional model updating method, because \color{black} they \color{black} may have a negligible impact on the accuracy of model updating.
				Hence, it is crucial to develop a method that quantifies uncertainties from limited data\color{black}, such as the case where \( N_\text{obs} = 1 \), \color{black} aimed at updating parameters related to such nonlinear seismic response.
		
		\subsection{Novel Likelihood Evaluation Method}\label{sec:novelmethod}
			The likelihood $p\left(\mathbf{X}_\text{obs} \middle| \boldsymbol{\uptheta}\right)$, as represented in Eq.~(\ref{bayes1}), is expressed \color{black} as: \color{black}
			\begin{equation}
				\label{eq:theta2xsim}
				p\left(\mathbf{X}_\text{obs} \middle| \boldsymbol{\uptheta}\right) = p\left(\mathbf{X}_\text{obs}\middle| \mathbf{X}_\text{sim}\right)
			\end{equation}
			Using an arbitrary multidimensional random variable $\boldsymbol{z}$, $p\left(\mathbf{X}_\text{obs} \middle| \mathbf{X}_\text{sim}\right)$ can be expressed as the marginalization of $p\left(\mathbf{X}_\text{obs}\middle|\boldsymbol{z}\right)$ \color{black} as: \color{black}
			\begin{equation}
				\label{dzzt}
				p\left(\mathbf{X}_\text{obs} \middle| \mathbf{X}_\text{sim}\right) =
				\int_{\mathcal{Z}}{p\left(\mathbf{X}_\text{obs}\middle|\boldsymbol{z}\right)p\left(\boldsymbol{z}\middle|\mathbf{X}_\text{sim}\right)}\>d\boldsymbol{z}
			\end{equation}
			where $\boldsymbol{z}=\left[z_1, z_2, \dots, z_{z_{dim}}\right]^T \in \mathcal{Z}$.
			It is assumed that
			$\mathbf{X}_\text{obs}$ and $\mathbf{X}_\text{sim}$ are conditionally independent given
			$\boldsymbol{z}$.
			Because $\boldsymbol{z}$ is a non-unique random variable, it is advantageous to set a manageable random variable $\boldsymbol{z}$ among the possible as discussed later.
			
			\color{black} Transforming \color{black} $p\left(\mathbf{X}_\text{obs}\middle|\boldsymbol{z}\right)$ using Bayes' theorem and Eq.~(\ref{eq:theta2xsim}), the likelihood $p\left(\mathbf{X}_\text{obs} \middle| \boldsymbol{\uptheta}\right)$ is expressed \color{black} as: \color{black}
			\begin{equation}
				\label{pzdzt}
				p\left(\mathbf{X}_\text{obs} \middle| \boldsymbol{\uptheta}\right) =
				c
				\int_{\mathcal{Z}}\frac{p\left(\boldsymbol{z}\middle|\mathbf{X}_\text{obs}\right)\>p\left(\boldsymbol{z}\middle|\mathbf{X}_\text{sim}\right)}{p\left(\boldsymbol{z}\right)}\>d\boldsymbol{z}
			\end{equation}
			In Eq.~(\ref{pzdzt}), $p(\mathbf{X}_\text{obs})$ is a constant and is denoted as $c$.
			If the probability distribution of $\boldsymbol{z}$ and the two conditional probabilities of $\boldsymbol{z}$ are obtained, the likelihood $p\left(\mathbf{X}_\text{obs} \middle| \boldsymbol{\uptheta}\right)$ is obtained using Eq.~(\ref{pzdzt}).

			In this context, the actual probability density function, $p\left(\boldsymbol{z}\middle|\cdot \right)$, is intractable \cite{VAE}.
			A numerical approach can be used to obtain the value of the likelihood $p\left(\mathbf{X}_\text{obs} \middle| \boldsymbol{\uptheta}\right)$ by approximating $p(\cdot)$ by $q(\cdot)$, \color{black} given by: \color{black}
			\begin{equation}
				\label{qzdzt}
				p\left(\mathbf{X}_\text{obs} \middle| \boldsymbol{\uptheta}\right) \approx
				c\int_{\mathcal{Z}}\frac{q\left(\boldsymbol{z}\middle|\mathbf{X}_\text{obs}\right)\>q\left(\boldsymbol{z}\middle|\mathbf{X}_\text{sim}\right)}
				{q\left(\boldsymbol{z}\right)}\>d\boldsymbol{z}
			\end{equation}
			It is difficult to \color{black} determine \color{black} proportionality constant $c$ here, but for MCMC it is sufficient to know the proportional value of the likelihood.
			
			$\boldsymbol{z}$ is a non-unique random variable. If $\boldsymbol{z}$ follows a low-dimensional and manageable distribution, $\boldsymbol{z}$ and $\mathcal{Z}$ represent the latent variable and latent space, respectively. This latent space captures hidden features or structures in data and plays a crucial role in complex data analysis. Here, $\boldsymbol{z}$ is a key parameter for quantifying the similarity between the observed feature sample $\mathbf{X}_\text{obs}$ and the simulated feature sample $\mathbf{X}_\text{sim}$.
			The utilization of the latent space enables in depth analysis of data, and Eq.~(\ref{qzdzt}) expresses the complex relationships between data using these latent features.
		
		\subsection{Application of VAE for Likelihood Estimation}
			\color{black} As \color{black} a practical example, this section \color{black} describes \color{black} an approach that utilizes the encoder of VAE \cite{VAE} as a probabilistic model that generates \color{black} output \color{black} $q(\cdot)$.
			An overview of VAE is provided in Appendix \ref{app:vae}.
			Using $q_\phi(\cdot)$ as a substitute for $q(\cdot)$ in Eq.~(\ref{qzdzt}) enables the evaluation of the \color{black} likelihood, expressed as: \color{black}
			\begin{equation}
				\label{eq:lkpm}
				p(\mathbf{X}_\text{obs}|\boldsymbol{\uptheta}) \approx
				c \int_{\mathcal{Z}} \frac{q_\phi(\boldsymbol{z}|\mathbf{X}_\text{obs}) \> q_\phi(\boldsymbol{z}|\mathbf{X}_\text{sim})}{q_\phi(\boldsymbol{z})} \> d\boldsymbol{z}
			\end{equation}
			where \color{black} subscript \color{black} $\phi$ denotes the parameters of the neural network constituting the encoder.
			This study proposes utilizing the output from the VAE encoder $q_\phi(\cdot)$.
			\color{black} Variable \color{black} $\boldsymbol{z}$ represents the latent variable, the output from the encoder in the VAE. The output of the encoder corresponding to the observed feature sample, is articulated as $q_\phi(\boldsymbol{z}|\mathbf{X}_\text{obs})$.
			\color{black}  Each sample of $q_\phi(\boldsymbol{z}|\mathbf{X}_\text{sim})$ is obtained from response analysis using a model with candidate parameter $\boldsymbol{\uptheta}$. \color{black}
			The latent variable $\boldsymbol{z}$ of the VAE follows a low-dimensional, independent normal distribution, thereby simplifying the integration of Eq.~(\ref{eq:lkpm}). The analytical calculation of Eq.~(\ref{eq:lkpm}) is shown in Appendix \ref{app:cal_lh}.

			The evaluation of likelihood using the proposed method represents a different interpretation of probability from the distance-based approach.
			The distance-based approach, which differs from our proposed method in that it estimates probability directly from observations, is suitable in cases such as estimating linear response parameters in structure models, where there is a limited amount of data.
			Conversely, the proposed approach introduces a subjective probability obtained from the dataset as prior information.
			\color{black} With this characteristic, the approach is \color{black} effective for updating beliefs with limited observations and is applicable for parameters related to the nonlinear responses of nonlinear models.
			Hence, both methodologies, each with its unique probability implications, are applicable to diverse situations in model updating, \color{black} and thus address \color{black} different challenges and objectives.

\begin{figure}
				\centering
				\includegraphics[width=0.97\columnwidth]{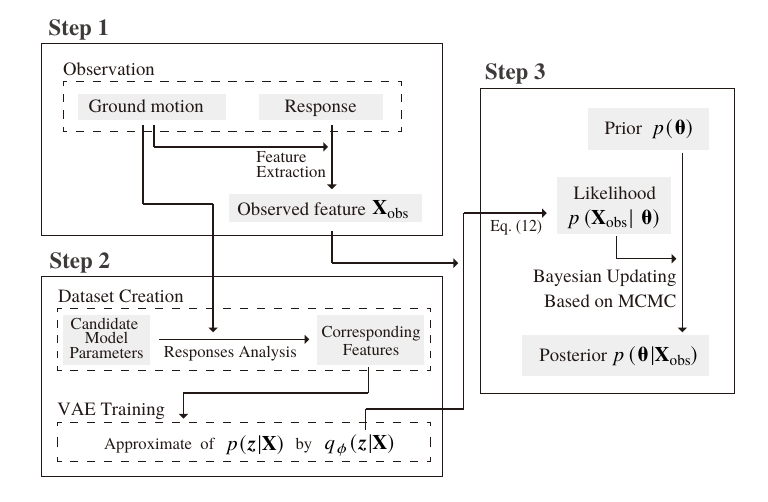}
				\caption{Schematic of the three-step model updating framework}
				\label{fig:flowchart}
			\end{figure}

			A framework incorporating the proposed likelihood evaluation technique based on MCMC algorithms is illustrated in Fig. \ref{fig:flowchart}.
			The framework consists of three steps.
			The first step involves extracting a feature from observational data.
			In this study, acceleration data can serve as observational data, and the frequency response function can serve as a feature.
			The second step involves training a VAE in the space of response analysis models. This is achieved by generating samples of various response analysis models with random uniformly distributed model parameters. Response analyses are then conducted using the observed ground motion as input, thereby creating a dataset for training and testing the VAE.
			In the third step, the model parameters are updated based on the observed data using MCMC to establish the posterior distribution.
			It is noteworthy that the likelihood evaluation of Eq.~(\ref{eq:lkpm}) is conducted using the trained VAE.

	\section{Quantification of Epistemic Uncertainty of MDOF Model Using Observation Variability}\label{sec:mdoftakeda}
		
		In this section, the model updating problem for the multi-degree-of-freedom (MDOF) model is examined to demonstrate the ability of the proposed approach in terms of quantifying the uncertainty resulting from the number of observation points \color{black} accounting for observation noise. In this numerical experiment, observation noise is imposed on both the response and input ground motion, wherein the true likelihood becomes intractable because of unknown true excitation. Calculating the likelihood is also challenging, even though the true input ground motion is known in this controlled setting, due to the high dimensionality of the data. Therefore, this example is particularly well-suited to the application of the approximate likelihood method.\color{black}
		
		\subsection{Analysis Model and Input Ground Motion }
\begin{figure}
				\centering
				\includegraphics[width=0.97\columnwidth]{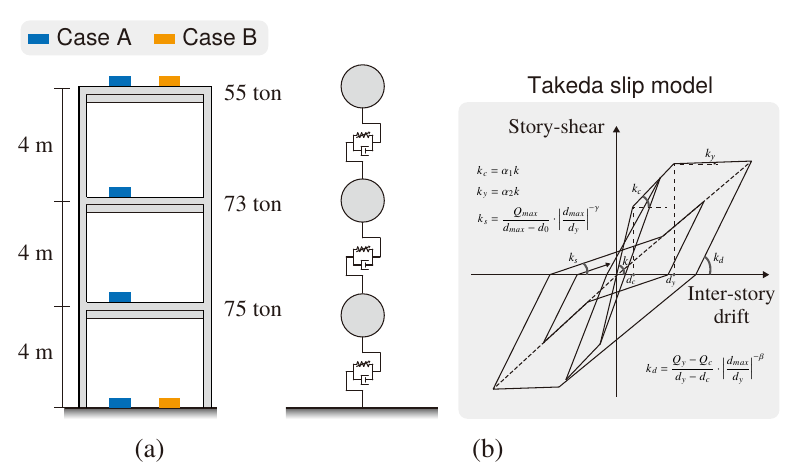}
				\caption{Three-degree-of-freedom system with the restoring force characteristics of the Takeda-slip model}
				\label{fig:model1}
			\end{figure}

			The validation cases adopted in the numerical experiments \color{black} presented \color{black} in this section and a schematic of the analytical model used are \color{black} depicted \color{black} in Fig.~\ref{fig:model1}.
			As shown in Fig.~\ref{fig:model1}(a), a three-story reinforced concrete (RC) structure is considered, and the validation \color{black} is divided \color{black} into two cases:
			Case A, where observations are conducted at each floor in addition to the input, and Case B, where observations are made solely at the top floor in addition to the input.
			As shown in Fig.~\ref{fig:model1}(b), a three degree of freedom model, which includes the restoring force characteristics of the modified Takeda-slip model \cite{takeda77}, was created to represent the target structure.

			The modified Takeda-slip model provides a comprehensive description of stiffness degradation (also known as slip) in the small drift region caused by the weakening of the column-beam joint anchorage performance.
			The model \color{black} was \color{black} developed based on the Takeda Model \cite{takeda70}, which explains the relationship between the displacement and restoring force of reinforced concrete structures.
			\color{black} There are seven model parameters, including \color{black} initial stiffness $k$, crack displacement $d_c$, yield displacement $d_y$, ratio of post-crack stiffness to initial stiffness $\alpha_1$, ratio of post-yield stiffness to initial stiffness $\alpha_2$, an index to determine the stiffness degradation rate under unloading $\beta$, and an index to define the slip behavior $\gamma$.
			The initial stiffness was assumed to vary between stories, whereas to reduce the number of unknown parameters, the other six parameters were assumed to be identical across all the stories. The number of parameters to be updated is nine ($n=9$).
			TABLE~\ref{tab:model1} presents the configuration of the assumed model parameters. The values of $k_1$, $k_2$, and $k_3$ in the table correspond with the initial stiffness of the first, second, and third story, respectively. Additionally, cracking and yielding were assumed \color{black} to \color{black} occur at 1/500 and 1/100 of the inter-story drift angle, respectively. Commonly used values are assigned to the other model parameters.
			
			\begin{table}
				\caption{Assumed model parameters of MDOF model}
				\label{tab:model1}
				\centering
				\small
				\renewcommand{\arraystretch}{1.25}
				\begin{tabular}{ccccccccc}
					\hline\hline
					$k_1$ (kN/mm) & $k_2$ (kN/mm) & $k_3$ (kN/mm) &
					$d_c$ (cm) & $d_y$ (cm) &
					$\alpha_1$ & $\alpha_2$&
					$\beta$ & $\gamma$\\
					\hline
					140 & 110 & 60 & 0.8 & 4 & 0.1 & 0.02 & 0.4 & 0.5 \\
					\hline\hline
				\end{tabular}
				\normalsize
			\end{table}
			
			The damping constant was set to 4\% for the first mode, and was assumed to be proportional to the instantaneous value of the stiffness.
			Although it is possible to assume the damping constant as one of the unknown model parameters, this study assumes that it is known. The restoring force characteristics affect more on the response at large amplitudes than the damping constant.
			When implementing the proposed method in actual structures, it is assumed that the damping constants are previously identified using observed data during minor earthquakes through system identification \cite{moesp}.
			
\begin{figure}
				\centering
				\includegraphics[width=0.97\columnwidth]{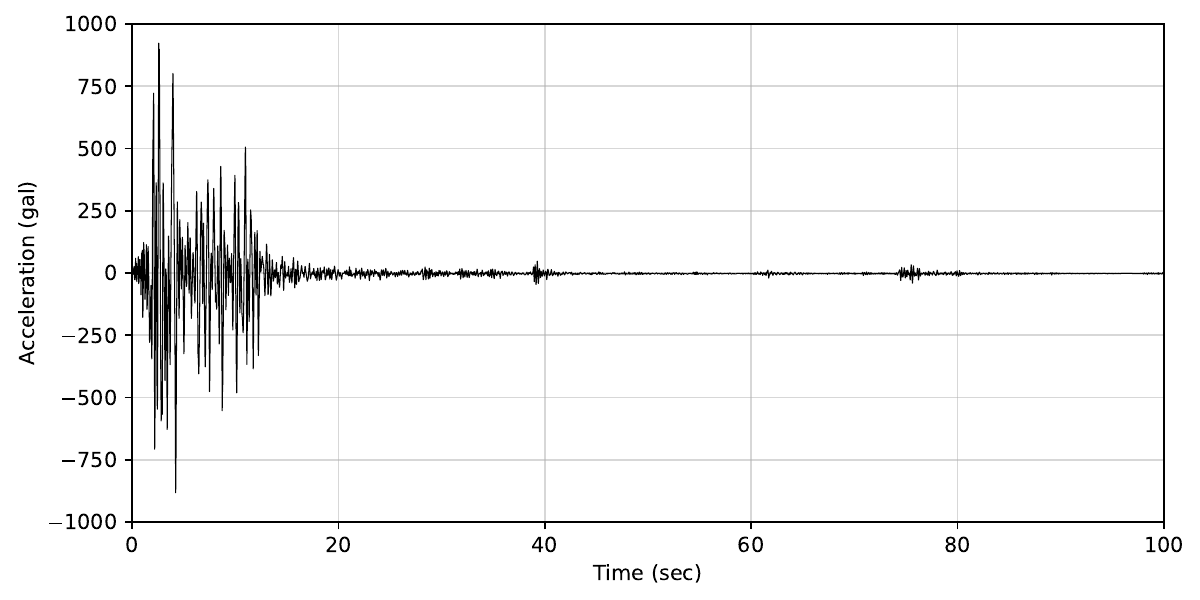}
				\caption{Waveform of the 2016 Kumamoto Earthquake, observed by the KiK-net strong-motion seismograph network (observation point: KMMH16, direction: E-W)}
				\label{fig:wave}
			\end{figure}

			In this study, a posterior distribution specific to the input ground motion \color{black} was obtained using \color{black} a single observation.
			The mainshock motion of the 2016 Kumamoto earthquake in Japan, observed at the KMMH16 station of KiK-net \cite{kiknet} in the E-W (East-West) direction with \color{black} the \color{black} sampling frequency of 100 Hz, is employed as the input ground motion.
			The duration of the motion is 100 s.
			The acceleration time history of ground motion utilized, as shown in Fig.~\ref{fig:wave}, \color{black} recorded the \color{black} peak acceleration of 922 gal, thereby meeting the appropriate conditions as an input ground motion of nonlinear response analysis.
			
			The seismic wave was upsampled to \color{black} the \color{black} frequency of 1000 Hz.
			The upsampled wave was then used in the response analysis, thereby resulting in the response wave of each floor.
			Further, the response waves were downsampled to 100 Hz.
			To introduce observational noise into the analysis, noise after a normal distribution with \color{black} the \color{black} mean of 0 and standard deviation of 0.1 gal was added to the input wave and to the response wave of each floor.
			Assuming that these noise-added waves were the ones observed, we proceeded to update the model parameters utilizing the proposed method.

			As discussed previously, our proposed method involves training the VAE each time after the observed records become available. Considering that significant seismic events are rare for each building, the approach using only a single observation is considered rational. Furthermore, it is important to underscore that the proposed model demonstrates robust performance across various ground motions, as detailed in \cite{Lee23ICASP} .
		
		\subsection{Dataset and Learning of VAE}
			The dataset used for training the VAE was created by generating uniform random numbers of the model parameters within the range \color{black} specified \color{black} in TABLE~\ref{tab:Dataset 1}.
			In this study, initial stiffness \color{black} was \color{black} set to a wide range that encompasses the true value. For application to an actual structure, a design value or a value obtained through system identification \cite{oku00} would be used in place of the true value.
			As presented in TABLE~\ref{tab:Dataset 1}, the cracking point (first degrading point) \color{black} was \color{black} reached within a range of inter-story drift angles of approximately 1/1600 to 1/200, while the yielding point \color{black} was \color{black} reached within \color{black} the \color{black} range of inter-story drift angles of 1/200 to 1/500. Further, the degrading ratio and slip index were set within a range that includes commonly used values. We set the upper and lower bounds of the model parameters in the same range for both Cases A and B, to discuss how the number of observation points affects the shape of the posterior distribution.
			
			The absolute acceleration of the response obtained from the response analysis was used to determine the frequency response function for the input acceleration, while the real and imaginary parts in the range of 0.1 to 5.24 Hz (512 points) were used as training data. \color{black} For the training dataset, 100,000 data were generated.
			Specifically, the dataset was \color{black} formed as a tensor with dimensions (100000, 6, 1, 512), wherein the numbers in the parentheses represent the number of data, the number of channels (corresponding with the real and imaginary parts of frequency response function at each floor), width, and length (corresponding with the frequency point of the frequency response function). In this format, the output size $m$ can be computed as 3072, which is the product of the number of channels, width, and length ($6 \times1\times512$).
			
			\begin{table}
				\caption{Ranges of model parameters for the MDOF model used in generating the training dataset}
				\label{tab:Dataset 1}
				\centering
				\small
				\renewcommand{\arraystretch}{1.25}
				\begin{tabular}{c|ccccccccc}
					\hline\hline
					&
					\multicolumn{1}{c}{$k_1$ (kN/mm)} & \multicolumn{1}{c}{$k_2$ (kN/mm)} & \multicolumn{1}{c}{$k_3$ (kN/mm)} &
					\multicolumn{1}{c}{$d_c$ (cm)} & \multicolumn{1}{c}{$d_y$ (cm)} &
					\multicolumn{1}{c}{$\alpha_1$} & \multicolumn{1}{c}{$\alpha_2$}&
					\multicolumn{1}{c}{$\beta$} & \multicolumn{1}{c}{$\gamma$}\\
					\hline
					upper bound & 200 & 160 & 120 & 2    & 8 & 0.25 & 0.05 & 1 & 1 \\
					lower bound & 100 & 60  & 20  & 0.25 & 2 & 0.05 & 0    & 0 & 0 \\
					\hline\hline
				\end{tabular}
				\normalsize
			\end{table}

\begin{figure}
				\centering
				\includegraphics[width=0.97\columnwidth]{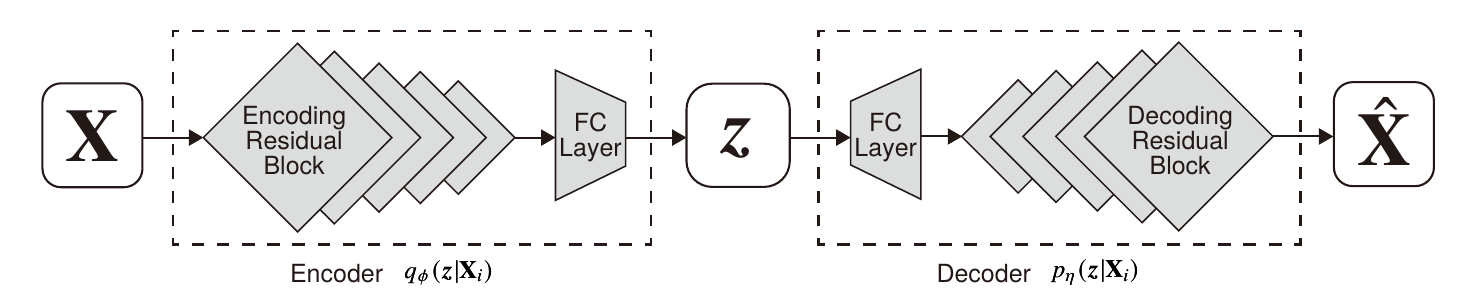}
				\caption{Architecture of the variational autoencoder (VAE) network for model updating}
				\label{fig:vaenet}
			\end{figure}

\begin{figure}
				\centering
				\includegraphics[width=0.97\linewidth]{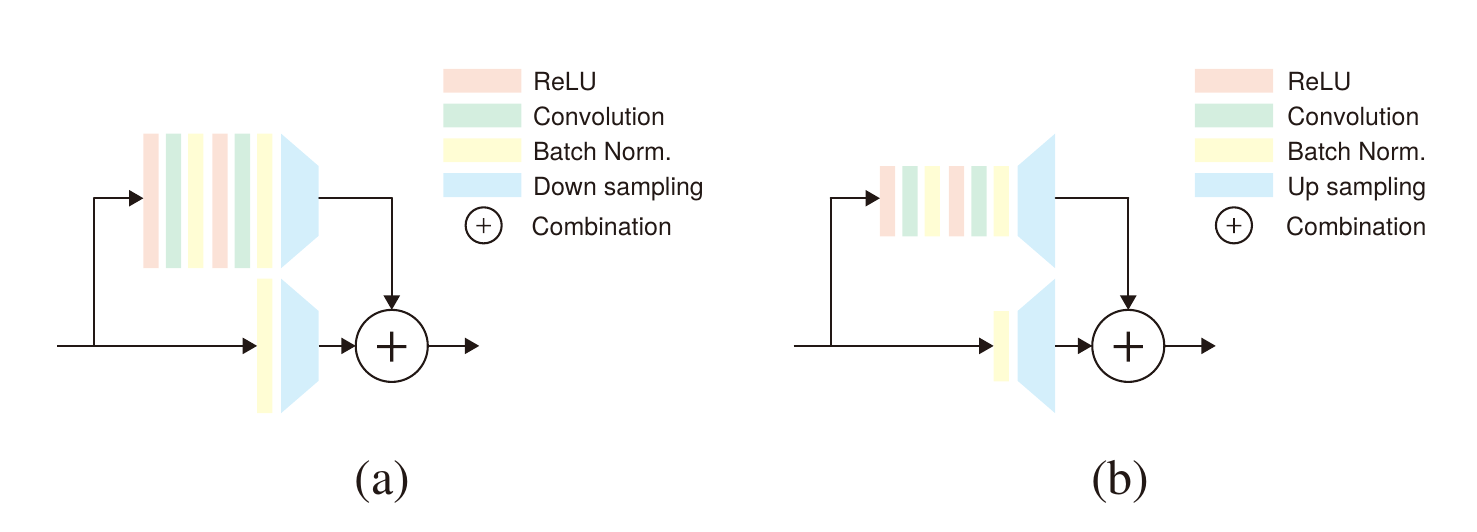}
				\caption{Residual blocks in the VAE: (a) encoding block, (b) decoding block}
				\label{fig:rb}
			\end{figure}

			The architecture of the VAE network used for model updating is shown in Fig. \ref{fig:vaenet}.
			The VAE comprises an encoder that transforms data $\mathbf{X}$ into a latent variable $\boldsymbol{z}$, and a decoder that transforms the latent variable $\boldsymbol{z}$ back into data $\hat{\mathbf{X}}$.
			The encoder has a structure that reduces the dimension of the input through a residual block \cite{resnet} \color{black} comprising \color{black} convolutional neural network (CNN) layers and fully connected (FC) layers, while taking the frequency response function as input.
			The decoder has a symmetrical structure to the encoder, thereby extending the input dimension of a residual block and FC layers, and using the latent variable as input.
			The residual blocks for the encoder and decoder are shown in Fig. \ref{fig:rb}.
			The encoding residual blocks shown in (a) expand the number of channels while downsampling to reduce the data length.
			Conversely, the decoding residual blocks shown in (b) reduce the number of channels and increase the length of the data \color{black} via \color{black} upsampling.
			In this study, we set the number of dimensions of the latent variable $\boldsymbol{z}$ to 10, which is slightly more than the nine model parameters.
			\color{black}
			The appropriate dimension of latent variable $\boldsymbol{z}$ ($\boldsymbol{z}_{dim}$) is a crucial factor throughout the training stage of the VAE and subsequent model updates.
			Setting $\boldsymbol{z}_{dim}$ too small increases uncertainty in updated model parameters due to the limited capability of VAE, whereas setting $\boldsymbol{z}_{dim}$ too large does not affect the result but results in unnecessary computational load.
			Therefore, it is important to find an adequately large $\boldsymbol{z}_{dim}$ that ensures sufficient model updating performance while maintaining a manageable computational load.
			This balance should be determined by monitoring the reconstruction performance during the training process.
			\color{black}
			
			The VAE was trained to minimize the loss function (Eq.~(\ref{eq:-E}) of Appendix~\ref{app:vae}) using the Adam optimiser \cite{adam} with \color{black} the \color{black} learning rate of 0.0001.
			We set the batch size to 64 to account for memory capacity, and stopped training after 1,000 epochs while confirming the decrease in the VAE loss function to ensure effective learning.
		
		\subsection{Bayesian updating with MCMC}
			The sampling of the model parameters based on the posterior distribution is performed using the Replica Exchange Monte Carlo method \cite{Swendsen86,remc96}, wherein a sample of each replica is generated based on the Metropolis-Hasting method \cite{metropolis53,Hastings70}.
			We set the likelihood function for each replica as follows:
			\begin{equation}
				\label{eq:like_temp}
				{}_{i}\mathcal{LH}\left({}_i^{} \boldsymbol{\vartheta}\right) \coloneq \left\{p\left(\mathbf{X}_\text{obs}|{ }_i^{} \boldsymbol{\vartheta}\right)\right\}^{T_i}
			\end{equation}
			where ${ }_i^{} \boldsymbol{\vartheta}$ is a model parameter vector sample of the $i$-th replica and $T_i$ is the temperature parameter that determines the form of the likelihood function for the $i$-th replica. \color{black} Furthermore, in \color{black} this study, we established eight replicas with $T_i$ values of 1, 2, 4, 8, 16, 32, 64, and 128.
			The exchange of samples between the $i$-th replica and the $j$-th replica occurs with the probability $p_e(i, j)$, expressed as
			\begin{equation}
				\label{eq:pe}
				p_e(i, j) \coloneq \frac{{}_{i}\mathcal{LH}({}_j^{} \boldsymbol{\vartheta})\>{}_{j}\mathcal{LH}({}_i^{} \boldsymbol{\vartheta})}
				{{}_{i}\mathcal{LH}({}_i^{} \boldsymbol{\vartheta})\>{}_{j}\mathcal{LH}({}_i^{} \boldsymbol{\vartheta})}
			\end{equation}
			The exchange according to the acceptance criteria was repeated 1,000 times for every 100 samples to obtain \color{black} the \color{black} total of 100,000 samples. The proposed method initializes each replica so that the sample in the latent variable space closely aligns with the observed data.
			Setting the initial value in this way \color{black} accelerates convergence \color{black} and allows \color{black} a shorter \color{black} burn-in period to be set.
			To verify the validity of the likelihood evaluation using the proposed method, the prior distribution was set as a non-informative distribution.

			The first 10,000 samples were considered as burn-in and hence were removed.
			The samples were further thinned by retaining every 30th sample from the remaining samples, thereby resulting in \color{black} the \color{black} total of 3,000 samples.
		
		\subsection{Result of Model Updating}
\begin{figure}
				\centering
				\includegraphics[width=0.97\columnwidth]{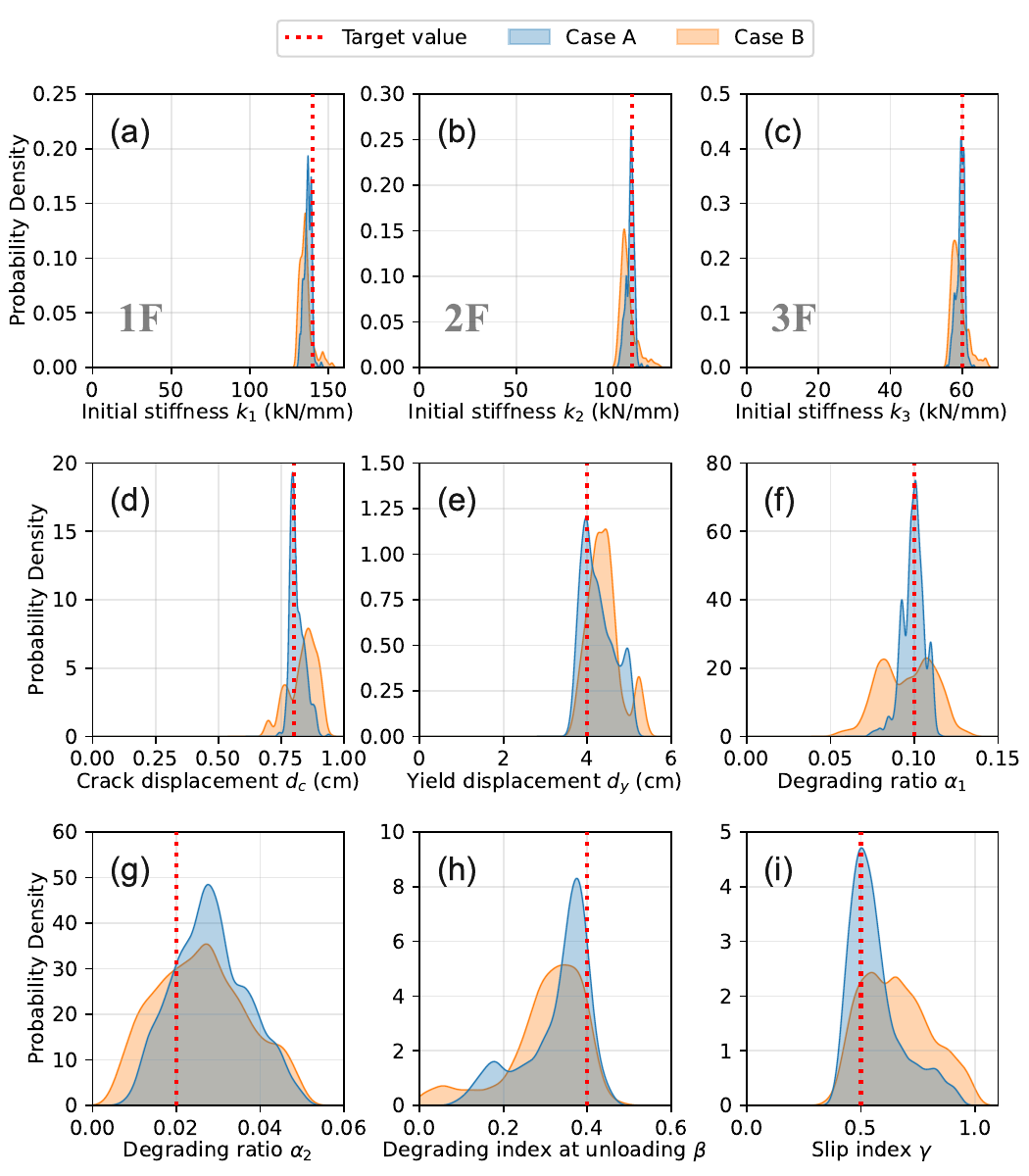}
				\caption{\color{black}Posterior distributions for model parameters in Cases A and B\color{black}}
				\label{fig:postdist}
			\end{figure}

			The shape of the posterior distribution, obtained \color{black} via \color{black} 1-dimensional Gaussian kernel density estimation, was determined using the 3,000 retained samples, as shown in Fig.~\ref{fig:postdist}.
			
			In the results of the proposed method \color{black} for \color{black} Cases A and B, it was observed that the posterior distributions for the nine model parameters consistently cover the true values.
			Narrow and sharp distributions were obtained for the initial stiffness, which are related to linear response and have a dominant impact on the response.
			However, parameters influencing only on the nonlinear response, such as stiffness reduction rates or slip index, had posterior distributions with a wider shape.
			In Case B, \color{black} with fewer \color{black} observations available, the distribution peaks were slightly shifted away from the true value \color{black} compared with those in \color{black}  Case A, where there are more observations available.
			In Case B, where the number of observations is smaller (uncertainty is larger), the breadths of the posterior distribution of most parameters were wider, and the peaks were lower than those of Case A.
			These results suggest that the proposed method not only facilitates the high-accuracy updating of multiple parameters related to complex nonlinear responses, but also quantitatively reflects uncertainty due to the lack of information originating from the number of observations.

			Using the 100 updated model samples selected randomly from retained MCMC samples and the true ground motion without noise, response analyses were conducted to estimate the overall response. The analysis parameters, including the damping ratio in the response analysis, were set as the same as those in the previous stages. \color{black} As a result, the \color{black} acceleration responses of each floor were obtained, as presented in Fig. 7.

			\begin{figure}
				\centering
				\includegraphics[width=0.97\columnwidth]{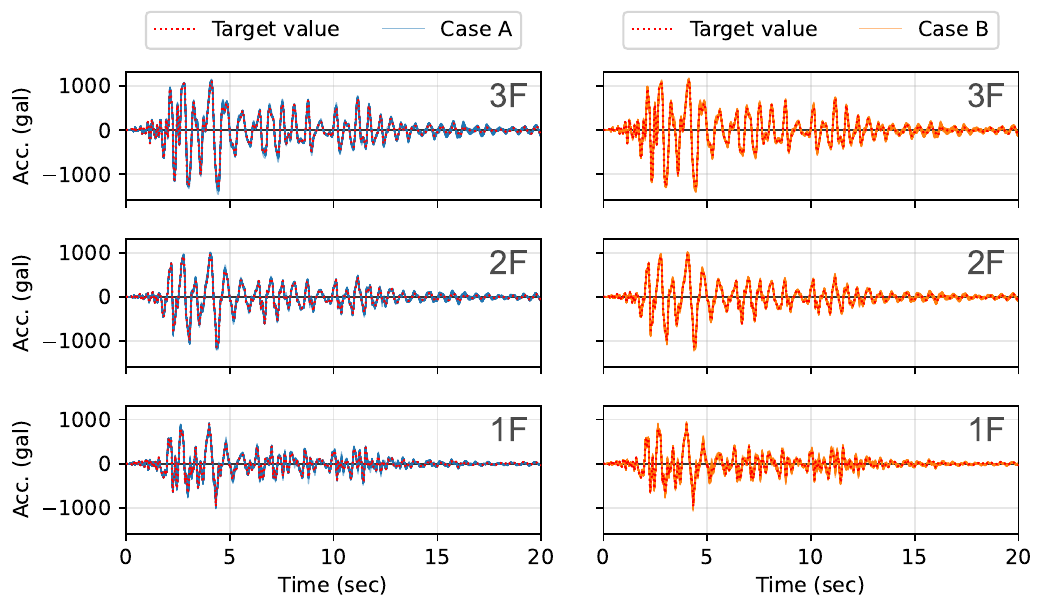}
				\caption{\color{black}Comparison of the response time histories between the updated models and the target model}
				\label{fig:respestm}
			\end{figure}

			Fig. 7 shows that the 100 response time histories using the updated models are almost identical for both Cases A and B.
			\color{black} Table \ref{tab:pfa} presents the means and standard deviations of each peak floor accelerations alongside the peak floor accelerations of the target model. Notably, in both Case A and Case B, variation between the time histories and peak floor accelerations are small, and the time history of the target model lies within the range of the 100 samples at every time step. \color{black}
			Considering that the posterior distributions of model parameters were wider, particularly in Case B, these results indicate that the proposed likelihood evaluation method effectively reflects parameter sensitivity in model updating.
			
			\begin{table}
				\color{black}
				\caption{\color{black}Comparison of the peak floor accelerations between the updated models and the target model}
				\label{tab:pfa}
				\centering
				\small
				\renewcommand{\arraystretch}{1.25}
				\begin{tabular}{c|c|cc|cc}
					\hline\hline
					& \multirow{2}{*}{Target value} & \multicolumn{2}{c|}{Case A} & \multicolumn{2}{c}{Case B} \\
					
					&       & Mean  & Standard deviation & Mean  & Standard deviation \\
					& (gal) & (gal) & (gal)              & (gal) & (gal)              \\
					\hline
					3F & 1273  & 1305  & 45.0               & 1293  & 51.1               \\
					2F & 1133  & 1133  & 28.0               & 1138  & 44.0               \\
					1F & 963   & 952   & 41.9               & 945   & 63.5               \\
					\hline\hline
				\end{tabular}
				\normalsize
			\end{table}

			\color{black}

	\section{Quantification of uncertainty due to lack of information on nonlinear response}\label{sec:sdofbilinear}
\begin{figure}
			\centering
			\includegraphics[width=0.7\columnwidth]{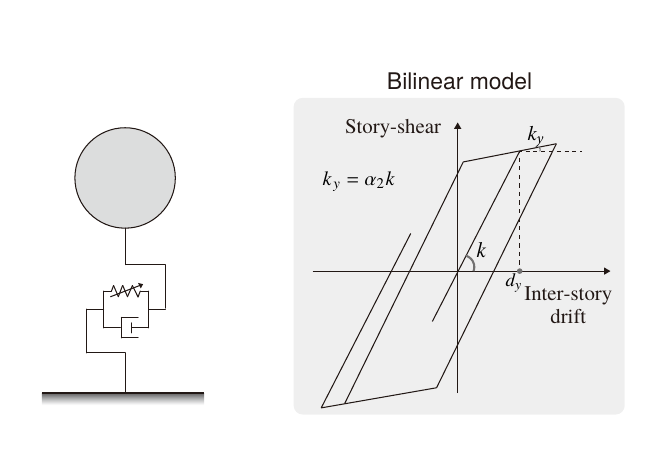}
			\caption{Single-degree-of-freedom system with the restoring force characteristics of bilinear model}
			\label{fig:model2}
		\end{figure}

		The relationship between the degree of nonlinearity of the response and breadths of the posterior distribution of the model parameters is explored using our proposed method.
		When severity of input ground motion decreases, information \color{black} regarding \color{black} the nonlinear response decrease from the observed records.
		For demonstration purposes, we \color{black} utilized \color{black} a single-degree-of-freedom shear spring model, as illustrated in Fig.~\ref{fig:model2}. \color{black} This model \color{black} is characterized by its bilinear restoring force and a relatively small number of parameters, \color{black} and thus serves \color{black} as an ideal example for validating our approach.
		
		\subsection{Target Analysis Model and Input Ground Motion}
			The model parameters comprise the natural frequency, which indicates the initial \color{black} stiffness; \color{black} the yield displacement, which indicates the displacement at the yield \color{black} point; \color{black} and the degrading ratio, which represents the post-yield stiffness relative to the initial stiffness ($n=3$).
			These three parameters \color{black} constitute \color{black} the bilinear restoring force characteristics.
			Here, the damping constant is assumed to be known, as in the previous example.
			The model parameters for the validation model group utilized in the numerical experiment are presented in TABLE~\ref{tab:vdmodel}.
			The group of validation models, comprising 45 models with \color{black} the \color{black} identical natural frequency of 2 Hz, parameterizes nine yield displacement ratios and five degrading ratios.
			The yield displacement ratio, expressed as a ratio to the maximum displacement obtained through linear response analysis, was used to measure the yield displacement.
			The response analyses were conducted by employing the identical input ground motion as in the previous example (see Fig.~\ref{fig:wave}), to acquire the waveform of response acceleration. The damping constant was set to 5\%, and the sampling period was upsampled to \color{black} the \color{black} interval of 0.001 s.
			The acceleration waveform obtained was presumed to \color{black} incorporate \color{black} observation noise (identical to \color{black} the \color{black} previous example), and the model parameters were updated using the proposed method.
			
			\begin{table}
				\caption{Assumed model parameters of validation models}
				\label{tab:vdmodel}
				\centering
				\small
				\renewcommand{\arraystretch}{1.25}
				\begin{tabular}{ccc}
					\hline\hline
					Natural frequency (Hz) & Yield displacement ratio                        & Degrading ratio           \\
					\hline
					2                      & 0.5, 0.55, 0.6, 0.65, 0.7, 0.75, 0.8, 0.85, 0.9 & 0.001, 0.1, 0.2, 0.3, 0.4 \\
					\hline\hline
				\end{tabular}
				\normalsize
			\end{table}
		
		\subsection{Dataset and Learning of VAE}
			The training dataset for the VAE was created using numerous response analysis models to obtain frequency response functions.
			To construct these response analysis models, we generated uniform random numbers of the model parameters within the ranges, as presented in TABLE~\ref{tab:Dataset2}.
			By configuring TABLE~\ref{tab:Dataset2}, half of the dataset comprises data that exhibits nonlinearity.
			The response analyses were conducted using the same input ground motion that was used for the target models. The damping constant is assumed to be known, following the same approach as in the previous example.
			The frequency response function was obtained relative to the input acceleration using the absolute acceleration of the response obtained from the response analysis.
			The real and imaginary components of the 512 dimensions within the frequency range of 0.1 to 5.22 Hz in the frequency response function were allocated to the corresponding channels in the training dataset.
			\color{black} The \color{black} total of 100,000 frequency response functions were generated for the training dataset. The dataset \color{black} was \color{black} formed as a tensor with dimensions (100000, 2, 1, 512), where the numbers in the parentheses represent the number of data, number of channels, width, and length. Using this format, the output size $m$ can be computed as 1024, which is the product of the number of channels, width, and length ($2 \times1\times512$).
			
			\begin{table}
				\caption{Ranges of model parameters for the
				SDOF model used in generating the training dataset}
				\label{tab:Dataset2}
				\centering
				\small
				\renewcommand{\arraystretch}{1.25}
				\begin{tabular}{c|ccc}
					\hline\hline
					Model parameter &
					Natural frequency (Hz) & Yield displacement ratio &
					Degrading ratio \\
					\hline
					upper bound & 5   & 1.8 & 0.5 \\
					lower bound & 0.5 & 0.2 & 0   \\
					\hline\hline
				\end{tabular}
				\normalsize
			\end{table}
			
			In this study, the employed VAE uses the same configuration as described in the previous example, including maintaining \color{black} the \color{black} latent variable dimension of 10, as illustrated in Fig.~\ref{fig:vaenet}.
			Based on the aforementioned procedures outlined, the training of the VAE was conducted using the Adam optimizer with \color{black} the \color{black} learning rate of 0.0001.
			Due to memory capacity considerations, \color{black} the batch size was set to 64. \color{black}
			The decision to train for 1000 epochs was based on monitoring the decrease in the loss function of VAE, which is further detailed in Appendix \ref{app:vae}.
		
		\subsection{Result of Model Updating}
			Consistent with the methodology outlined in the previous example, the MCMC simulation \color{black} detailed \color{black} in this chapter was conducted using the observed waveforms from the 45 validation models.
			The conditions of the MCMC simulation, including the selection of starting samples and the number of sampling iterations, were identical to those previously described.
			Samples were obtained based on the posterior distribution, thereby maintaining a similar procedural fidelity as established in the aforementioned section.
			Although visualising the posterior distributions of all 45 models is omitted for space reasons,
			the relatively simple model parameter updating problem facilitated very accurate estimations.
			
\begin{figure}
				\centering
				\includegraphics[width=0.75\columnwidth]{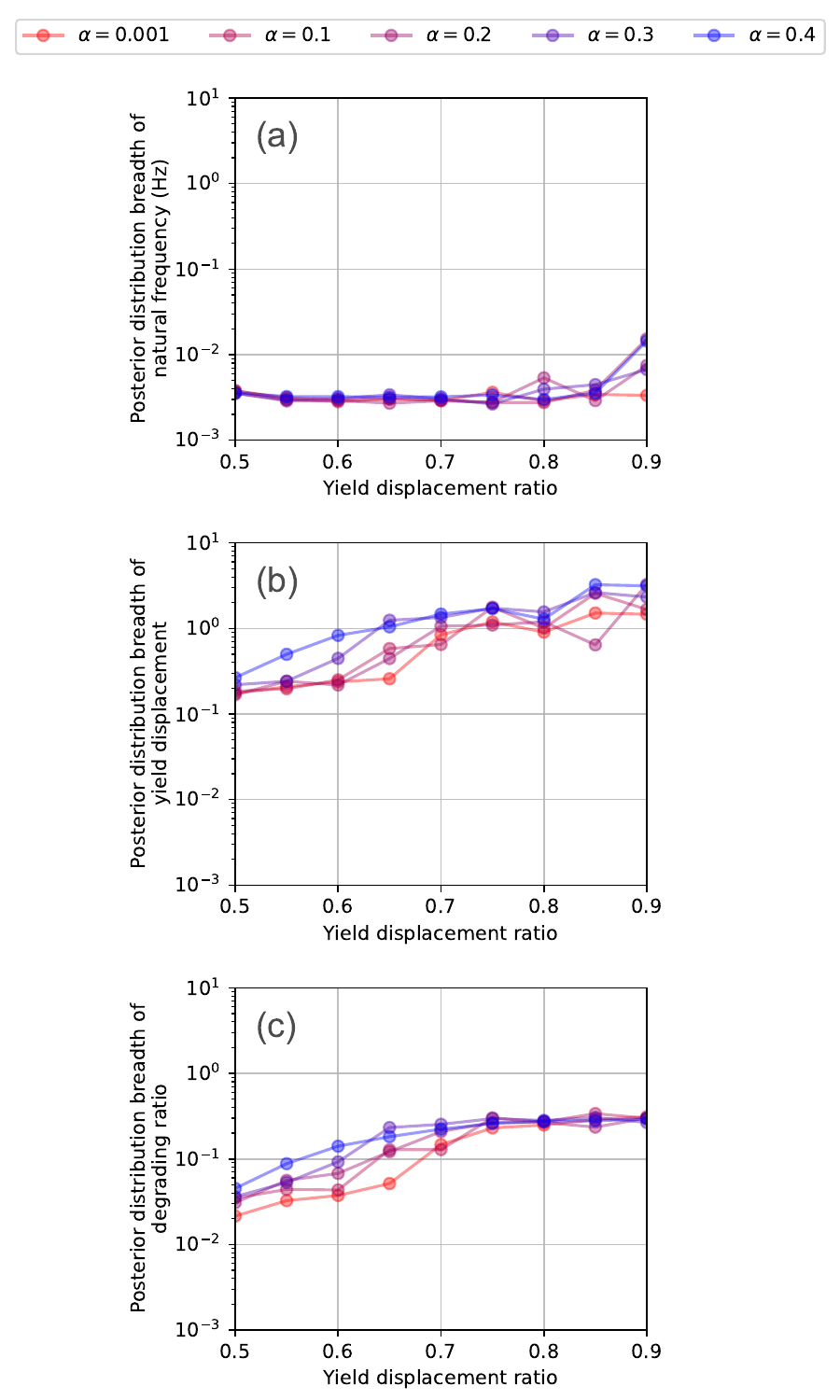}
				\caption{Variation of posterior distribution breadth for validation models}
				\label{fig:postwidth}
			\end{figure}

			For this validation, our main objective \color{black} was \color{black} to examine the trends in \color{black} the \color{black} uncertainty evaluation, particularly those attributed to limited information, without focusing on assessing the estimation accuracy.
			The breadth of the posterior distribution \color{black} was \color{black} determined as the sample standard deviation \color{black} using Eq.~(\ref{eq:width}) and \color{black} is \color{black} displayed against the yield displacement ratio \color{black} in \color{black} Fig.\ref{fig:postwidth}.
			
			\begin{equation}
				\label{eq:width}
				\sigma_i= \sqrt{\frac{1}{n_s}\sum_{j=1}^{n_s}{\left({ }_1^{} \vartheta_{i}^{(j)}- \Bar{{ }_1^{} \vartheta_{i}}\right)^2}}
			\end{equation}
			Here, the number of MCMC adopted samples is denoted as $n_s$, whereas ${ }_1^{} \vartheta_{i}^{(j)}$ represents the $i$-th model parameter of the $j$-th sample of the 1st replica.
			Additionally, $\Bar{{ }_1^{} \vartheta_{i}}$ represents the mean of the series ${ }_1^{} \vartheta_{i}^{(j)}$, \color{black} expressed as: \color{black}
			\begin{equation}
				\Bar{{ }_1^{} \vartheta_{i}}=
				\frac{1}{n_s}\sum_{j=1}^{n_s}{ }_1^{} \vartheta_{i}^{(j)}
			\end{equation}
			The values on the horizontal axis \color{black} in \color{black} Fig.~\ref{fig:postwidth} correspond with the target values of the yield displacement ratio, \color{black} while \color{black} the colours of the graph correspond with the target values of the degrading ratio $\alpha$.
			An increase in the prescribed values of the yield displacement ratio and degrading ratio results in a decrease in nonlinearity of response.
			This could be perceived as an increase in uncertainty due to insufficient data on the nonlinear response.
			The natural frequency, which draws information from the complete time domain of the response data, is indicative of a consistently negligible breadth of the posterior distribution throughout the range. Nevertheless, for yield displacement ratios above 0.8, the breadth of the posterior distribution of the natural frequency shows a slight increase. This trend is due to the correlation between the estimated natural frequency and the other two parameters associated with the nonlinear response. The large uncertainty in these two parameters in this range influences the uncertainty in the natural frequency.
			Conversely, the yield displacement ratio and degrading ratio, which obtain information from the nonlinear response, exhibited a tendency for the breadth of the posterior distribution to increase as the target values of the yield displacement ratio and degrading ratio increased.
			These findings show that the proposed technique accounts for the uncertainty arising from insufficient information on the nonlinear response and adjusts the breadth of the posterior distribution accordingly.
			
	\section{Conclusion}
		This study proposed a novel approximate Bayesian technique for updating the parameters of a nonlinear response analysis model. This technique considers the lack of information due to limited number of observations \color{black} through a VAE. \color{black} We conducted numerical experiments to investigate the shape of the posterior distribution of model parameters, thereby validating the proposed methodology.
		
		\color{black} For the \color{black} numerical experiments, a multi-degree-of-freedom system \color{black} was used \color{black} with a modified Takeda-slip hysteresis characteristics model, which exhibits relatively complex hysteresis features.
		\color{black} As a result, \color{black} the proposed approach successfully attained posterior distributions close to the target values. The shape of the posterior distribution depends on the density of observations.

		Subsequent experiments utilized a single-degree-of-freedom system with a bilinear hysteresis model, which exhibits relatively simple hysteresis features. The shape of the posterior distribution also depends on the degree of nonlinearity in observed records. These results show that the proposed method can evaluate likelihood effectively, hereby considering uncertainty due to limited information.
		
		This study introduced a rigorous procedure for calculating likelihood, employing MCMC with response analysis, which requires relatively large computational resources. Developing a methodology to reduce such computational resources will be the focus of future research.
		\color{black}
		Additionally, future work will include applying the proposed method to stochastic model updating in \color{black} a multiple-observation \color{black} scenario \color{black} as well as \color{black} disentangling the estimation of observation noise and parameter distributions. This approach will \color{black} thus enable \color{black} examination of the effects of different noise levels on posterior distributions, \color{black} which \color{black} were not addressed in this study.

	\section{Data Availability Statement}
		All data, models, or code that support the findings of this study are available from the corresponding author upon reasonable request.
		\appendix

	\section{Overview of Variational Auto-encoder}\label{app:vae}
\begin{figure}
			\centering
			\includegraphics[width=0.95\columnwidth]{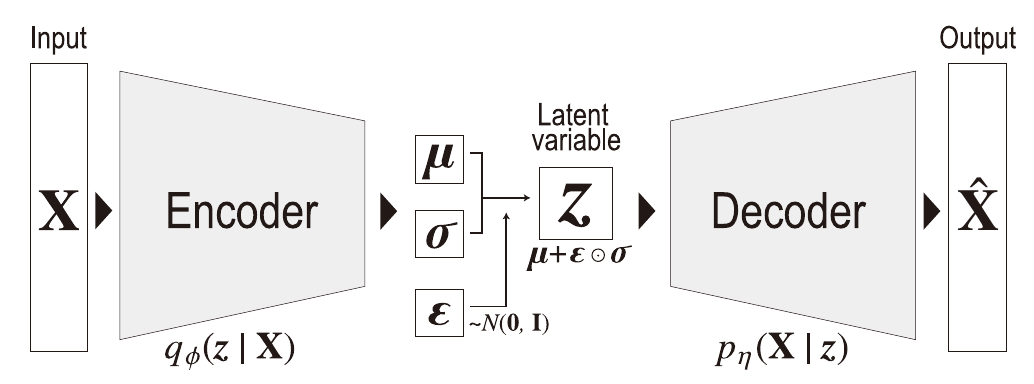}
			\caption{VAE architecture}
			\label{fig:vae}
		\end{figure}

		The proposed method utilizes variational auto-encoders (VAEs) \cite{VAE}. These are generative models that aim to replicate the distribution of a training dataset from a latent variable. An architecture of a VAE, as illustrated in Fig.~\ref{fig:vae}, comprises two main components: an encoder and a decoder. The encoder, a neural network, transforms the input data $\mathbf{X}$ into a latent variable $\boldsymbol{z}$. The decoder, another neural network, uses this latent variable to reconstruct data, $\hat{\mathbf{X}}$, that reconstruct the original input in dimensions.

		Although VAEs are similar to conventional autoencoders \cite{AE}, VAEs treat the outputs of the encoder and decoder as random variables individually. This is achieved by what is known as the reparameterization trick. In this approach, the standard deviation of output from the encoder $\boldsymbol{\sigma}$ is multiplied with a random number $\boldsymbol{\varepsilon}$, drawn from the multi-variate standard normal distribution. Later, the mean $\boldsymbol{\mu}$ is added to generate the random vector $\boldsymbol{z}$. The reparameterization trick enables the backpropagation of gradients through random nodes, thereby facilitating the training of the neural network.

		The decoder in a VAE aims to optimize the marginal likelihood $p_\eta(\mathbf{X})$ reproducing the input data $\mathbf{\mathbf{X}}$.
		Due to the intractability of $p_\eta(\mathbf{X}|\boldsymbol{z})$, the output of the encoder $q_\phi(\boldsymbol{z}|\mathbf{X})$, which approximates the likelihood of $\boldsymbol{z}$ for a given $\mathbf{X}$, is utilized instead. Here, $\phi$ and $\eta$ represent the parameters of the encoder and decoder neural networks, respectively.

		For each training instance $\mathbf{X}_i$, the output of the decoder $p_\eta(\mathbf{X}_i)$ is independent of $\boldsymbol{z}$ and follows the distribution $q_\phi(\boldsymbol{z}|\mathbf{X}_i)$. This relationship is expressed in Eq.~(\ref{eq:logpx}):
		\begin{equation}
			\label{eq:logpx}
			\log{p_\eta}\left(\mathbf{X}_i\right) = \mathcal{L}\left(\eta,\phi,\mathbf{X}_i\right) + D_{KL}\left(q_\phi\left(\boldsymbol{z}\middle|\mathbf{X}_i\right)||p_\eta\left(\boldsymbol{z}\middle| \mathbf{X}_i\right)\right) \geq \mathcal{L}\left(\eta,\phi,\mathbf{X}_i\right)
		\end{equation}
		where $D_{KL}$ denotes the Kullback-Leibler divergence. In the context of VAEs, an error function is defined as expressed in Eq.~(\ref{eq:elbo}). The neural network parameters $\phi$ and $\eta$ are fine-tuned to minimize this error function.
		
		\begin{equation}
			\label{eq:elbo}
			-\mathcal{L}\left(\eta,\phi,\mathbf{X}_i\right) = -\mathbb{E}_{q_\phi\left(\boldsymbol{z}\middle|\mathbf{X}_i\right)}\left[\log{p_\eta}\left(\mathbf{X}_i\middle|\boldsymbol{z}\right)\right] + D_{KL}\left(q_\phi\left(\boldsymbol{z}\middle|\mathbf{X}_i\right)||p_\eta(\boldsymbol{z})\right)
		\end{equation}
		In \color{black} the above \color{black} equation, $\mathbb{E}_{q_\phi\left(\boldsymbol{z}\middle|\mathbf{X}_i\right)}\left[\cdot\right]$ represents the expected value operation for the variable $\boldsymbol{z}$.
		
		The first term on the right-hand side of Eq.~(\ref{eq:elbo}) represents the reconstruction error, with the aim to align the output of the decoder with the input data $\mathbf{X}_i$. This alignment is approximated using the Monte Carlo method, wherein the random number $\boldsymbol{z}_{i,l}$ drawn from $q_\phi\left(\boldsymbol{z}\middle|\mathbf{X}_i\right)$ is employed, \color{black} as: \color{black}
		\begin{equation}
			\label{eq:-E}
			-\mathbb{E}_{q_\phi\left(\boldsymbol{z}\middle|\mathbf{X}_i\right)}\left[\log{p_\eta}\left(\mathbf{X}_i\middle|\boldsymbol{z}\right)\right] \approx -\frac{1}{L}\sum_{\boldsymbol{z}_{i,l}}{\log{p_\eta}\left(\mathbf{X}_i|\boldsymbol{z}_{i,l}\right)}
		\end{equation}
		where, $L$ represents the number of samples used in the Monte Carlo estimation. In mini-batch learning, wherein training data is segmented into smaller batches for processing, \color{black} $L=1$ is generally set \color{black} when the batch size (indicating the data count in each batch) is adequately large (as suggested in the original paper, it is 100).
		To compute $\log{p_\eta}(\mathbf{X}_i|\boldsymbol{z}_{i,l})$, \color{black} a loss function must be defined \color{black} for the output data of the decoder. Cross-entropy loss is used for discrete datasets, whereas mean square error is preferred for continuous data.
		In this study, we \color{black} employed \color{black} the mean square error due to the nature of our continuous data.
		It is important to note that the underlying assumption of employing the mean square loss function is that the conditional probability density $p_\eta(\mathbf{X}_i|\boldsymbol{z}_{i,l})$ is multivariate Gaussian.

		The second term in Eq.~(\ref{eq:elbo}) acts as a regularization factor, thereby ensuring \color{black} that \color{black} the output of the encoder $q_\phi\left(\boldsymbol{z}\middle|\mathbf{X}_i\right)$ closely approximates $p_\eta(\boldsymbol{z})$, modeled as a standard normal distribution. \color{black} As \color{black} the encoder \color{black} is trained \color{black} to transform the input data $\mathbf{X}$ into independent standard normal variable $\boldsymbol{z}$ in a nonlinear manner, the likelihood evaluation method \color{black} can be \color{black} established. It is important to note that this assumption does not rely on previously observed data or model parameter distributions, but rather defines the learning process of VAEs.

	\section{Analytical Method for Likelihood Computation}\label{app:cal_lh}
		
		In this appendix, we detail the analytical method for computing likelihoods.
		As the latent variable $\boldsymbol{z}$ is trained to conform to a multi-dimensional standard normal distribution, we decompose and integrate each dimension as delineated below:
		\begin{equation}
			\int_{\mathcal{Z}} \frac{q_\phi(\boldsymbol{z}|\mathbf{X}_\text{obs}) \> q_\phi(\boldsymbol{z}|\mathbf{X}_\text{sim})}{q_\phi(\boldsymbol{z})} \> d\boldsymbol{z}
			= \prod_{i=1}^{z_{dim}} \int_{-\infty}^{\infty} \frac{q_\phi(z_i|\mathbf{X}_\text{obs}) \> q_\phi(z_i|\mathbf{X}_\text{sim})}{q_\phi(z_i)} \> dz_i
		\end{equation}
		
		The integration over each dimension of the latent variable is further examined, as expressed in Eq.~(\ref{eq3:qphi2}):
		\begin{equation}
			\label{eq3:qphi2}
			\int_{-\infty}^{\infty} \frac{q_\phi(z_i|\mathbf{X}_\text{obs}) \> q_\phi(z_i|\mathbf{X}_\text{sim})}{q_\phi(z_i)} \> dz_i
			\eqcolon L_{z_i}
		\end{equation}
		where $ q_\phi\left(z_i \mid\mathbf{X}_\text{obs}\right) $ and $ q_\phi\left(z_i \mid \mathbf{X}_\text{sim}\right) $ represent encoder outputs and are thereby modeled as normal distributions via the reparameterization trick. The term $ q_\phi\left(z_i\right) $ approximates a normal distribution due to the regularization error component.
		Assuming $z_{i{|\mathbf{X}_\text{obs}}}\sim N({}_1^{}\mu_i^{},{}_1^{}\sigma_i^{})$, $z_{i{|\mathbf{X}_\text{sim}}}\sim N({}_2^{}\mu_i^{},{}_2^{}\sigma_i^{})$, $z_{i}\sim N({}_3^{}\mu_i^{},{}_3^{}\sigma_i^{})$, the term $L_{z_i}$ can be expressed \color{black} as: \color{black}
		\begin{equation}
			\label{eq3:qphi3}
			L_{z_i}
			= \sqrt{\frac{{}_3^{}\sigma_i^2}{2\pi{}_1^{}\sigma_i^2{}_2^{}\sigma_i^2}}\int_{-\infty}^{\infty}e^{-\frac{\left(z_i-{}_1^{}\mu_i^{}\right)^2}{2{}_1^{}\sigma_i^2}-\frac{\left(z_i-{}_2^{}\mu_i^{}\right)^2}{2{}_2^{}\sigma_i^2}+\frac{\left(z_i-{}_3^{}\mu_i^{}\right)^2}{2{}_3^{}\sigma_i^2}}\>dz_i
		\end{equation}
		As expressed in Eq.~(\ref{eq3:qphi3}), the exponential portion becomes a quadratic polynomial, further simplifying the expression as in Eq.~(\ref{eq3:qphi4}).
		\begin{equation}
			\label{eq3:qphi4}
			L_{z_i}=d_i\int_{-\infty}^{\infty}{e^{-\left(a_iz_i^2+b_i z_i + c_i\right)}dz_i}
		\end{equation}
		Regarding the coefficients, we have $a_i=\frac{1}{2{}_1^{}\sigma_i^2}+\frac{1}{2{}_2^{}\sigma_i^2}-\frac{1}{2{}_3^{}\sigma_i^2}$, $b_i=-\frac{{}_1^{}\mu_i^{}}{{}_1^{}\sigma_i^2}-\frac{{}_2^{}\mu_i^{}}{{}_2^{}\sigma_i^2}+\frac{{}_3^{}\mu_i^{}}{{}_3^{}\sigma_i^2}$, $c_i=\frac{{}_1^{}\mu_i^2}{2{}_1^{}\sigma_i^2}+\frac{{}_2^{}\mu_i^2}{2{}_2^{}\sigma_i^2}-\frac{{}_3^{}\mu_i^2}{2{}_3^{}\sigma_i^2}$, $d_i=\sqrt{\frac{{}_3^{}\sigma_i^2}{2\pi{}_1^{}\sigma_i^2{}_2^{}\sigma_i^2}}$.
		As ${}_1^{}\sigma_i^{}, {}_2^{}\sigma_i^{} \ll {}_3^{}\sigma_i^{}$, it follows that $a_i > 0$.
		
		The equation transforms as follows:
		\begin{equation}
			\label{eq3:qphi6}
			L_{z_i}=d_ie^\frac{b_i^2-4a_ic_i}{4a_i}\sqrt{\frac{\pi}{a_i}}\int_{-\infty}^{\infty}{\sqrt{\frac{a_i}{\pi}}e^{-a_i\left(z_i+\frac{b_i}{2a_i}\right)^2}dz_i}
		\end{equation}
		The integrand within the integral represents a normal distribution with mean $ -\frac{b_i}{2a_i} $ and standard deviation $ \sqrt{\frac{1}{2a_i}} $, culminating in the result \color{black} in \color{black} Eq.~(\ref{eq3:qphi7}).
		\begin{equation}
			\label{eq3:qphi7}
			L_{z_i}=d_ie^\frac{b_i^2-4a_ic_i}{4a_i}\sqrt{\frac{\pi}{a_i}}
		\end{equation}
		
		The process, as summarized in Eq.~(\ref{eq3:qphi8}), involves breaking down the complex integrals into more manageable expressions for each dimension of $\boldsymbol{z}$.
		\begin{equation}
			\label{eq3:qphi8}
			\int_{\mathcal{Z}} \frac{q_\phi(\boldsymbol{z}|\mathbf{X}_\text{obs}) \> q_\phi(\boldsymbol{z}|\mathbf{X}_\text{sim})}{q_\phi(\boldsymbol{z})} \> d\boldsymbol{z}
			= \prod_{i=1}^{z_{dim}} d_ie^\frac{b_i^2-4a_ic_i}{4a_i}\sqrt{\frac{\pi}{a_i}}
		\end{equation}

		\bibliography{ascexmpl-new}

\end{document}